\documentclass[10pt,conference,compsocconf,letterpaper]{IEEEtran}
\makeatletter
\def\ps@headings{%
\def\@oddhead{\mbox{}\scriptsize\rightmark \hfil \thepage}%
\def\@evenhead{\scriptsize\thepage \hfil \leftmark\mbox{}}%
\def\@oddfoot{}%
\def\@evenfoot{}}
\makeatother
\usepackage{color}
\usepackage{soul}
\setulcolor{red}
\sethlcolor{yellow}

\def\squareforqed{\hbox{\rlap{$\sqcap$}$\sqcup$}}
\def\qed{\ifmmode\squareforqed\else{\unskip\nobreak\hfil
\penalty50\hskip1em\null\nobreak\hfil\squareforqed
\parfillskip=0pt\finalhyphendemerits=0\endgraf}\fi}

\usepackage{graphicx, amssymb, subfigure,setspace}
\usepackage{times}
\usepackage{algorithm}
\usepackage{algorithmicx}
\usepackage{algpseudocode}
\usepackage{amsmath}
\usepackage{makecell}
\usepackage{tabularx}
\usepackage{textcomp}

\usepackage[long]{optional}

\usepackage{comment}
\newtheorem{lemma}{Lemma}
\newtheorem{theorem}{Theorem}

\begin{document}

\title{vSkyConf: Cloud-assisted Multi-party \\Mobile Video Conferencing}
%or Meeting Room in the Cloud: Cloud-assisted Peer-to-Peer Mobile Video Conferencing

\author{\IEEEauthorblockN{
Yu Wu\IEEEauthorrefmark{1}, Chuan Wu\IEEEauthorrefmark{1}, Bo Li\IEEEauthorrefmark{2}, Francis C.M. Lau\IEEEauthorrefmark{1}}
\IEEEauthorblockA{
\IEEEauthorrefmark{1}Department of Computer Science, The University of Hong Kong, Email: \{ywu,cwu,fcmlau\}@cs.hku.hk}
\IEEEauthorblockA{ \IEEEauthorrefmark{2}Dept.~of Computer Science and Engineering, Hong Kong University of Science and Technology,
Email: bli@cse.ust.hk}
}

\maketitle
\begin{abstract}
As an important application in today's busy world, mobile video conferencing facilitates people's virtual face-to-face communication with friends, families and colleagues, via their mobile devices on the move. However, how to provision high-quality, %battery conservative,
 multi-party video conferencing experiences over mobile devices is still an open challenge. Our survey on 7 representative mobile video conferencing applications shows that at most $2-4$ concurrent participants can be supported in one conference. %xx applications have stringent requirements on xxx streaming rates that the mobile device should support ({\bf Give the numbers}), and a xxx bps video conference of xxx applications can run for at most xxx minutes on an iPhone 4S. 
 The fundamental problem behind is still a lack of computation and communication capacities on the mobile devices, to scale to large conferencing sessions. In this paper, we present {\em vSkyConf}, a cloud-assisted mobile video conferencing system, to fundamentally improve the quality and scale of multi-party mobile video conferencing. By novelly employing a surrogate virtual machine in the cloud for each mobile user, we allow fully scalable communication among the conference participants via their surrogates, rather than directly. The surrogates exchange conferencing streams among each other, transcode the streams to the most appropriate bit rates, and buffer the streams for the most efficient delivery to the mobile recipients. %For efficient video conferencing streams broadcast among participants, 
 A fully decentralized, optimal algorithm is designed to decide the best paths of streams and the most suitable surrogates for video transcoding along the paths, such that the limited bandwidth is fully utilized to deliver streams of the highest possible quality to the mobile recipients. We also carefully tailor a buffering mechanism on each surrogate to cooperate with optimal stream distribution. Together they guarantee bounded, small end-to-end latencies and smooth stream playback at the mobile devices, in the video conferencing sessions.
% All these designs free the mobile devices from power-consuming processing and communication, %fully exploit the benefits of mobile cloud computing, and renders a highly scalable, energy efficient, high-quality mobile video conferencing solution.
 We have implemented {\em vSkyConf} based on Amazon EC2 and verified the excellent performance of our design, as compared to the widely adopted unicast solutions.
\end{abstract}

\section{Introduction}
\label{sec:introduction}

%Confronted with the soaring oil prices, over the past few years, a wealth of video conferencing systems, facilitating convenient on-line collaborations among separate parties instead of expensive transportations, have been widely deployed to drive a clean energy economy. Advances of today in the digital signal processors (DSP) technologies enable more mobile embedded devices to exploit the ever efficient processing capabilities and hence open up new visions into evolutions of phone calls, as part of every day life for ordinary people. 

%The challenges are further exacerbated when it comes to mobile-based video conferencing, due to the constrained processor capabilities, limited battery life, unstable network bandwidths and heterogeneous configurations, compared to its counterpart, {\em i.e.}, traditional room-based or PC-based video conferencing.

Online video conferencing has been widely deployed for virtual, face-to-face communication among separate parties, as a greener solution to replace many of the energy-expensive conference travels. Advances in mobile and wireless communication technologies have enabled mobile users to exploit new evolution of phone calls --- mobile video conferencing calls --- as part of their everyday life, anytime anywhere on the move.

{\small
\begin{table}
	\centering
	\caption{Representative Mobile Video Conferencing Apps: a Comparison}
				\begin{tabular}{|p{1.5cm}|p{1.5cm}|p{1.9cm}|p{2cm}|}%{c|c|c|c}
					\hline
						 \textbf{App} & \textbf{Structure} & \textbf{Max. \# of Participants} & \textbf{Cellular Call Support} \\
					\hline
						 FaceTime & P2P & 2 & no \\
					\hline
						 LifeSize & S/C & 4 & yes \\
					\hline
						 Skype & P2P & 2 & yes\\
					\hline
						 Vidyo & S/C & 4 & yes\\
					\hline
						 Fring & P2P & 4 & yes\\
					\hline
						 Fuze & S/C & 4 & yes\\
					\hline
						 Tango & P2P & 2 & yes \\
					\hline
				\end{tabular}
				\label{tab:apps}
				\vspace{-6mm}
\end{table}
}

A number of mobile video conferencing applications have emerged \cite{skype}\cite{lifesize}\cite{tango}\cite{vidyo}%({\bf add citations})
. Many rely on expensive, dedicated architectures, {\em e.g.}, multiple control units (MCU), to process signaling messages, transcode ingress session streams and disseminate multiple streams to each end device. Such a centralized solution is limited in scalability, and the expensive up-front investment prohibits its wide adoption by small or medium institutions, let alone individual users. Distributed, peer-to-peer (P2P) based mobile video conferencing solutions have also been deployed, {\em e.g.}, Skype mobile \cite{skype}, which leverages intermediate super nodes for session relays. 

Can the existing mobile video conferencing systems support high-quality, multi-party video conferencing over mobile devices? We seek the answer by conducting a survey of $7$ representative applications, with results given in Table \ref{tab:apps}.
%We observe that cellular support has been achieved as a consensus by competitors, since FaceTime will also allow cellular video calls with the coming iOS6 upgrade. 
We observe that applications with infrastructure support (S/C) tend to support more concurrent users under expensive user subscription fees \cite{lifesize}\cite{vidyo}, %({\bf Chuan: add citation})
 while P2P-based solutions are reluctant to allow group video calls, for a fear of compromising call qualities. Skype is believed to provide the most decent call quality, but it only supports two-way visual communication on mobile phones (while web-based Skype allows $10$ concurrent communication sessions among premium user accounts), and so is Tango \cite{tango}%({\bf Chuan: add citation})
 . Most applications stick to one streaming rate; Skype and Fring recently declare Dynamic Video Quality (DVQ) by adapting video bit rates according to the transient connection conditions \cite{fring}%({\bf Chuan: add citation})
 , but only a limited number of bit rates are supported, {\em e.g.}, medium quality streams of $256$ kbps and higher quality streams of $512$ kbps in Skype. %Fring is the only pioneering app supporting multi-party video chats and deserves praises for its striving ahead of the curves,  despite the degraded qualities.

%We observe that only xx applications can support xxx concurrent participants in one conference, xx applications have stringent requirements on xxx streaming rates that the mobile device should support ({\bf Give the numbers}), and a xxx bps video conference of xxx applications can run for at most xxx minutes on an iPhone 4S. 

Based on this state of the art, we conclude that high-quality, multi-party mobile video conferencing is still a pending goal to achieve. We summarize the key challenges as follows: (1) The workload on each node in a video conferencing session, in terms of both processing and transmission, scales quadratically to the size of the session, which makes it challenging to use mobile devices for multi-party video conferencing. % That is probably the reason why most distributed video conferencing systems only allow two-party video chats, while voice-only chats are feasible for multi-party conferencing. %Another hurdle lies in the strict time sensitivity. To some extent as packet losses are acceptable, the perceived call quality drops significantly once a small time delay is violated.
 (2) Mobile users are equipped with different devices and downlink speeds; a high-quality solution should %``Universal fairness'' should be achieved by granting 
 enable differentiated call qualities to different users, instead of a homogeneous video broadcast quality enforced by the low-end users, as in a traditional solution. %Besides, codec-agnostic features are well appreciated by converting the streams into compatible ones playable at different devices.    %{\bf Explain the reasons leading to this challenge, like the discussions in (1)}

%(3) The never-long-enough battery life. The limited power capacity significantly limits the processing and transmission capacities of individual mobile devices, preventing them from participating in a fully distributed multi-party conferencing solution. 

%(4) delay/smoothness of the playback

In this paper, we present {\em vSkyConf}, a cloud-assisted mobile video conferencing system, to fundamentally enable high-quality,
 multi-party video conferencing over heterogeneous mobile devices. The cloud computing paradigm offers ubiquitously accessible computing resources, with on-demand resource provisioning at modest cost. The paradigm particularly compensates well for the inherent resource deficiencies of mobile devices, and catalyzes the undergoing evolution in the burgeoning mobile computing industry. In {\em vSkyConf}, we dynamically provision a virtual machine in the cloud as the exclusive surrogate for a dialed-in mobile user. Each mobile device uploads its stream to its surrogate and downloads others' streams from the surrogate; the surrogates exchange conferencing streams among each other, transcode the streams to the most appropriate bit rates, and buffer the streams for the most efficient delivery to the mobile recipients. By leveraging the more powerful processing capabilities and stable wired network bandwidths, mobile users shift those otherwise on-device tasks to the cloud, yielding superior power reduction and quality enhancement, as well as achieving fully scalable communication among the conference participants.

To realize such a design, several key questions remain to be answered: (1) How should we map the quadratically increasing session flows to the links between surrogates, to achieve satisfactory streaming experience with latency guarantees? (2) To which surrogates should the necessary transcoding tasks be assigned, considering different computing capacities of the surrogates? (3) In dynamic networking environments where jitters happen frequently, how should a surrogate help to smooth out jitters for mobile devices? %(4) Frequent reconstruction of the dissemination paths will inversely incur more jitters and hence affect the overall system performance, then in what event adjustments be taken to make the system adaptive enough? (5) Last but not least, how to better improve real-time streaming transmission qualities over unreliable networks, since tradition TCP/IP stack does not suit well? ({\bf elaborate the questions when the main part is done.})

To address these issue, a fully decentralized, optimal algorithm is designed to decide the best paths of streams and the most suitable surrogate for video transcoding along the paths, such that bandwidth capacities in the system are fully utilized to deliver streams of the highest possible quality to the mobile recipients. We also carefully tailor a buffering mechanism on each surrogate to cooperate with optimal stream distribution. Together they guarantee bounded, small end-to-end latencies and smooth stream playback at the mobile devices, in the video conferencing sessions.
%Our design achieves the following goals: {\em Highly scalable}, {\em Battery efficient}, {\em High streaming quality, low delay}. Though targeted at video conferencing systems, our proposed framework applies well in other time-stringent resource-intensive rate-diversified multi-user streaming applications. 
 We have implemented {\em vSkyConf} based on Amazon EC2. Experiments in the real-world settings reveal the high scalability, full adaptability, and excellent video conferencing qualities achieved by our design, as compared to the widely adopted unicast solutions.

%Cloud computing, offering the promises of ubiquitous accessible computing resources with its abundant on-demand provisioning agilities and pay-as-you-go pricing models, has proven its strengths in various fields and incubated more-than-ever-imagined new types of real-life applications, at modest costs. The new paradigms particularly compensates well for the inherent resource deficiencies of mobile devices, and catalyze the undergoing evolutions in burgeoning mobile computing industries. The promising combinations of cloud system and mobile devices are believed to realize a new ad hoc fashioned video conferencing system, on which multiple friends can simultaneously communicate visually on the move.

%We set to explore integration of a generic cloud system and the fast-growing smartphone technologies by designing and implementing a cloud-assisted multi-party video conferencing system on top of the best-effort IP network. In lieu of costly and dedicated central infrastructure, {\em e.g.}, MCU, our system dynamically provision virtual machines in the cloud as the exclusive surrogate for dialled-in mobile users. By leveraging more powerful processing capabilities and stable wired network bandwidths, mobile users shift those otherwise on-device tasks, {\em e.g.}, transcoding and call disseminations, from the smartphones to the cloud, yielding stunning power consumption and quality enhancements. 

%Our contributions in this study are as follows.

%{\em First},propose a model...

%{\em Second},routing...

%{\em Third} dynamic jitter design...

The remainder of this paper is organized as follows. We conduct a thorough literature survey in Sec.~\ref{sec:relatedwork}. Unique challenges and the system architecture are presented in Sec.~\ref{sec:arch}. Design details unfold in Sec.~\ref{sec:design}, followed by Sec.~\ref{sec:experiments} introducing the deployed prototype as well as real-world evaluations. Finally, Sec.~\ref{sec:conclusions} concludes the paper.%({\bf Chuan: complete})
\section{Related Work}
\label{sec:relatedwork}

Despite extensive studies during the past decades, video conferencing (VC) has recaptured people's interest in this new ``Smartphone'' era, with a series of works and systems springing up recently \cite{vidyo}\cite{DBLP:journals/tmm/PonecSCLC11}\cite{Chen:2011:CTL:1989240.1989270}\cite{DeCicco:2008:SVR:1496046.1496065}\cite{DBLP:conf/infocom/HuangMLW11}\cite{DBLP:journals/ton/LiangZL11}, which can be categorized into Server-to-Client (S/C) based and Peer-to-Peer (P2P) based solutions.

The network-layer solution to naturally support VC is still IP  multicast \cite{Ratnasamy:2006:RIM:1151659.1159917}. However, its weakness of scalability, difficulties of deployment and security issues still prohibit it from being a practical choice.

Cloud computing, as a natural agile solution, compensates well for the deficiencies of mobile devices for media streaming, in terms of both processing and bandwidth supports. Traditional players \cite{vidyo}\cite{webex} in the VC marketplace have recently claimed their support to mobile users of different platforms via their private clouds. WebEx \cite{webex} builds up their services using Cisco clouds. Vidyo \cite{vidyo} even advertises the slogan ``Conferencing-as-a-Service'', and offers a complete solution by provisioning virtual MCUs on top of their VidyoRouters \cite{vidyo}, bearing similar flavors to their traditional dedicated infrastructures. In contrast to such centralized solutions for enterprise users, our work novelly provisions a VM surrogate for each ordinary mobile user in an IaaS cloud, in a more affordable manner. %little %Although it appears straightforward to leverage the agile cloud services to compensate for deficiencies of mobile devices in a mobile-oriented multi-party video conferencing. Although it appears straightforward to harness the agile cloud services to compensate for deficiencies of mobile devices in a mobile-oriented multi-party video conferencing system, not that much efforts have been witnessed in academia for such a centralized solution,  due to expensive investments veering ordinary

%The advances of mobile technologies have turned more and more things which would have been practical only years ago commonplace.

Another series of work try to exploit scalable video coding (SVC) to enable differentiated services to users with different available bandwidths. Huang {\em et al.}~\cite{DBLP:conf/infocom/HuangMLW11} leverage clouds to encode videos into layered rates, but the encoding complexities inevitably incur intolerable delays for a time-sensitive application like video conferencing. Besides, the output bit rates for SVC encoders are restricted within a range, not flexible under a much more dynamic network condition.

Compared to the S/C model, P2P is deemed as a more promising structure. Both Ponec {\em et al.}~\cite{DBLP:journals/tmm/PonecSCLC11} and Chen {\em et al.}~\cite{Chen:2011:CTL:1989240.1989270} formulate utility maximization problems and enable multi-party VC by building multi-rate multicast trees. They focus more on the streaming rate allocation over physical links, but do not investigate much the transcoding flexibilities. %while adopting simple delay guarantees by restricting the tree depths. However, our real-life experiments show that the number of hops is not directly related to the delays perceived by end users.
 Liang {\em et al.}~\cite{DBLP:journals/ton/LiangZL11} leverage the upload capacities of ``helpers'' from other swarms, in similar ways as adopted by Skype (not Skype mobile). Though promising, it is difficult to achieve in cases of mobile users who are reluctant to contribute resources to strangers, due to constrained batteries and expensive cellular data fees. De Cicco {\em et al.}~\cite{DeCicco:2008:SVR:1496046.1496065} conduct solid measurements evaluating Skype video-rate adaptabilities to bandwidth variations and reveal only $450$ kbps can be achieved even under a good network condition. Those parameters can act as good references in our evaluation of the {\em vSkyConf} prototype implementation.

The dominant solution in most existing P2P-based mobile VC applications is still pair-wise unicast, {\em e.g.}, Fring \cite{fring}, Tango \cite{tango}, etc., due to simplicity of implementation. However, the limited uplink bandwidths of mobile devices lead to a constrained swarm size. Our real-life experiments in Sec.~\ref{sec:experiments} also reveal its susceptibility to network jitters especially for long-haul sessions, compared with {\em vSkyConf}.

A recent work by Feng {\em et al.}~\cite{yuan12} leverages inter-datacenter networks to maximize the overall throughput of all conferencing sessions, based on intra-session network coding. {\em vSkyConf} considers both dynamic session routing and adaptive session transcoding, and advocates to exploit a cloud infrastructure for mobile video conferencing. %well as a novel application layer protocol suite.
 Little effort has been devoted to building a cloud-assisted multi-party mobile video conferencing system, catering for the needs of ordinary mobile users in their daily life. {\em vSkyconf} is designed with this goal in mind. The framework and associated protocol suite can also apply to other delay-sensitive, rate-differentiated, multi-party mobile video streaming applications.

%The flaw of the network-coding solution, latency issue. intra-session network coding. Though promising,

%network layer solution: IP multicast.

%\cite{DBLP:journals/tmm/PonecSCLC11}: utility maximization based rate control optimal peer upload bandwidth usage, multi-rate multicast tree. primal-dual solution. Targeting on a small-swarm sized video conferencing as ours. focus more on rates instead of the transcoding efforts. tree packing. 

%\cite{Chen:2011:CTL:1989240.1989270} assumes the capacity network can be any where.

%ad hoc simulcast approach, pair-wise communications.

%Peer-to-Peer:
%collaborative usage of upload bandwidth of peers.

%\cite{Akkus:2011:PMV:1889383.1889430} applying a layered codec, requires each peer's bandwidth  can send and receive one full-quality video stream, which is a hard constraint for mobile users.

\section{Architecture and Design Objectives}
\label{sec:arch}

In this section, we highlight the key components and design principles of the cloud-assisted video conferencing system, {\em vSkyConf},  with detailed designs unveiled in Sec.~\ref{sec:design}. 

\subsection{Architecture and Key Modules}

\begin{figure}[!t]
	\begin {center}
	\includegraphics[width=0.45\textwidth]{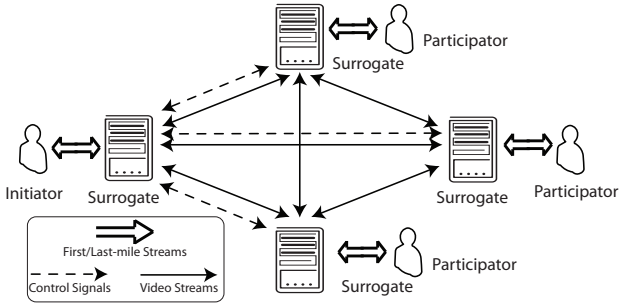}
	\vspace{-2mm}
	\caption{The architecture of vSkyConf.}
	\label{fig:arch}
	\end {center}
	\vspace{-6mm}
\end{figure}

{\em vSkyConf} enables efficient, peer-to-peer fashioned, multi-party mobile video conferencing via an IaaS cloud, with the architecture presented in Fig.~\ref{fig:arch} %({\bf Chuan: change `Intra-cloud Streams' to `Video Streams'})
. We refer to a video conference call among multiple mobile users as a {\em session}. The user which starts the conference call is the {\em initiator} of the session. Each user in a session produces a video stream, via the camera on its mobile device, sends the stream to other users, as well as receives streams produced by all the other users.

A surrogate, {\em i.e.}, a virtual machine (VM) instance, is created in the IaaS cloud for each mobile user. The IaaS cloud consists of disparate data centers in different geographic locations, and the surrogate for each mobile user is assigned on a data center proximate to the user. As a proxy for the mobile device, a mobile user's surrogate is responsible for the following: (i) session maintenance, by exchanging control messages with other surrogates in a timely and efficient manner; (ii) video dissemination and transcoding, by receiving the video stream its mobile user produces, transcoding it into appropriate format(s), distributing it to its own and other users' surrogates, and the other way round as well; (ii) efficient video buffering for its mobile user, for timely, smooth and robust streaming to the corresponding device. A mobile user just needs to send the stream it generates and receive streams others produce to and from its surrogate, and is effectively freed from power-consuming processing and communication. A gateway server in {\em vSkyConf} loosely keeps track of participating users and their surrogates, which can be implemented by a standalone server or VMs in the IaaS cloud.

% A surrogate is only visible to the user to whom it is assigned and is responsible to send and receive the video flows for that specific user, but also help transcode and forward the video streams to other surrogates in the same video conferencing session, without users' interactions. 

\begin{figure}[!t]
	\begin {center}
	\includegraphics[width=0.45\textwidth]{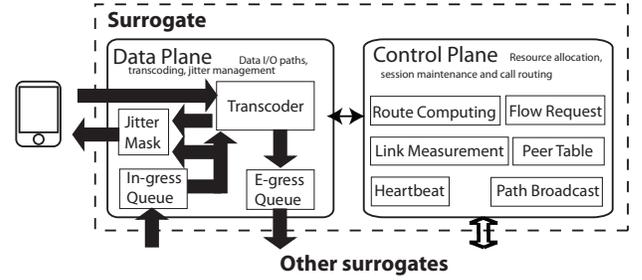}
	\vspace{-2mm}
	\caption{The key modules of a surrogate.}
	\label{fig:blocks}
	\end {center}
	\vspace{-6mm}
\end{figure}

The key modules implemented on a single surrogate is depicted in Fig.~\ref{fig:blocks}, which can be divided into two parts: the {\em control plane} and the {\em data plane}. % and the {\em user interface}. ({\bf Chuan: enclose the modules belonging to a surrogate in a dotted rectangular shape and mark `Surrogate'; put `Other surrogates' besides the arrows on the left; transrater $\rightarrow$ transcoding? jitter manage $\rightarrow$ `jitter management'; add an arrow directly leading from `In-gress Queue' to `Jitter mask'; `Resource allocations'$\rightarrow$`Resource allocation'; remove `QoS' block; how about add an arrow from `Jitter Buffer' to the mobile phone icon, and an arrow from the mobile phone to `Webcam', and change `Webcam' to `Input Buffer from User'})

{\em Control Plane} is the brain of the surrogate, responsible for control signaling between this surrogate and neighboring surrogates. It measures the latencies and bandwidths on the connections from/to neighboring surrogates, and all the collected information is stored in the ``peer table'', which constructs a partial view of the video conferencing topology from this surrogate's point of view. Utilizing the collected information, the surrogate computes routing paths for streams from its corresponding mobile user to other mobile users, and participates in the construction of optimal video dissemination trees. It also monitors the call qualities and determines the best video encoding parameters (codecs, bitrates, etc.) for streams from/to its mobile device.

{\em Data Plane} is responsible for processing in/out video streams, in terms of both transcoding and forwarding, as directed by the control plane. The video stream from its mobile user is captured continuously and disseminated to other surrogates after necessary transcoding. In the reverse direction, all video streams from other mobile users, via their respective surrogates, are transcoded into appropriate rates (if necessary)  and delivered to the mobile user by a key module ``jitter mask'', which deals with random jitters caused by fluctuations of processing and network latencies, as well as any anomalies along the dissemination paths.

%{\em User Interface} is a thin layer between the surrogate and the mobile user. The mobile device sends its captured stream to the surrogate via the interface, and downloads video streams from other mobile users through the jitter buffer in the interface. The mobile device's hardware configurations and the user's explicit commands can also be %collected and
 %communicated to the control plane for route calculations and transcoding decisions, via this interface layer.

\subsection{Design Objectives}

Our design of {\em vSkyConf} observes the following principles.

{\bf Decentralized Control}. Except necessary bootstrapping from the gateway server, {\em vSkyConf} aims to rely as little as possible on the central control, for session maintenance and route computation. Each session is to be maintained by the surrogate of the initiator of a conference session, in order to provide good scalability and flexibility. The video routing and transcoding decisions are to be made in a fully distributed fashion by collaborations among surrogates. Considering the mobile users can join and leave the system dynamically, fail-over mechanisms are also smartly integrated, to guarantee robustness of each session.

{\bf Self-Evolving Routing Topology with Full Adaptivity}.
A best routing topology for disseminating the stream from each participant should be built among the surrogates in a conference session, which achieves a small end-to-end latency to each of the other users and fully exploits the available bandwidths among the surrogates. Transcoding decisions to convert the original stream to acceptable formats/bit rates of the recipients should be made at the best point along the dissemination paths, according to different computation capacities of the surrogates and needs of downstream mobile devices. The routing paths and transcoding points should be dynamically evolving, according to the current bandwidth and latency among surrogates and wireless connectivity to the mobile users. {\em vSkyConf} proposes a dynamic routing and transcoding algorithm to achieve these objectives.

{\bf Synchronized Playback}.
One primary knotty issue in a multi-party video call is to keep all the streams played synchronously, without any noticeable lagging or leading streams. A straightforward approach to offsetting skewness among multiple streams as also applied by {\em vSkyConf} is to impose a latency, forcing the leading streams to wait for the lagging ones. That, however, leads to another associated synchronization problem in a different dimension, due to different ``wall clocks'' at different mobile users. Traditional solutions seek help from Network Time Protocol (NTP) \cite{ntp} servers to adjust each user's system time. In contrast, we seek to design a different simple measure to allow all the participatory surrogates to calibrate skewness of their own clocks against that of the session initiator.

{\bf Robust, Smooth Video Streaming}.
In a practical distributed system, various random events may happen. For instance, upstream surrogates in a dissemination path may suddenly drop offline or may over-claim their link bandwidths and latencies due to measurement errors, etc. To guarantee smooth stream playback at each mobile user even in cases of inaccurate route computation, {\em vSkyConf} designs an advanced error correction mechanism to search for better call routing paths before the call quality drops, by monitoring a carefully designed jitter buffer with pre-configured thresholds.

\section{Detailed Design}
\label{sec:design}
We present detailed design of {\em vSkyConf}, to achieve the design principles presented in Sec.~\ref{sec:arch}.

\subsection{Session Maintenance}
\label{sec:signaling_proto}

%Traditional video conferencing systems usually adopt Session Initiation Protocol (SIP) \cite{} for session establishment. But SIP is more appropriate for either a centralized solution wherein proxy servers or registrar servers help establish the session among the user agent clients (UAC), or a flat peer-to-peer fashion under complex multi-rounded negotiations between UACs. In {\em vSkyConf}, mobile users don't bother to negotiate directly between each other, but only hand over the tasks to their surrogates. Therefore, we design a proof-of-concept protocol suite in {\em vSkyConf}, where all the signal communications are realized in a gossip style between neighbouring surrogates in the same conferencing session. Functionally, the protocol suite can be categorized into four groups.

\begin{figure}[h]
	\begin {center}
	\includegraphics[width=0.45\textwidth]{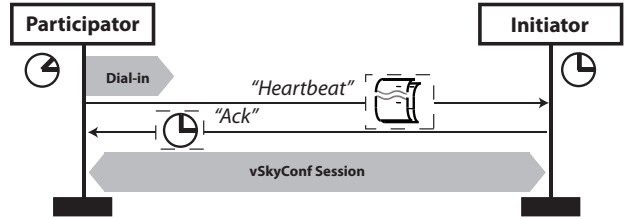}
	\caption{Clock synchronization among different surrogates.}
	\label{fig:clockskewness}
	\end {center}
	\vspace{-6mm}
\end{figure}

{\em Establishment}: When a mobile user logs in to the {\em vSkyConf} system via the gateway server, it is assigned a surrogate VM. The gateway can maintain information on a pool of available, pre-initiated VMs in the IaaS cloud, and assign one from the pool to a mobile user based on geographic proximity of the two, to expedite the service. The surrogate of the session initiator finds out IP addresses of surrogates of the other online users from the gateway server, which it wishes to invite to join a video conferencing session. The initiator then relies no more on the gateway server: it contacts and invites the interested participants through their surrogates directly, and maintains a list of IP addresses of all active surrogates in the session. Each participant sends periodical ``heartbeat'' messages to the session initiator, and receives the time-stamped ``ack'' from the initiator which is used to calibrate the local ``clock'' skewness against the initiator's, as shown in Fig.~\ref{fig:clockskewness}. The updated lists of IP addresses are periodically broadcast to all active participant surrogates from the initiator as well. In this way, the load on the gateway server is significantly alleviated by initiators of different conferencing sessions, and one gateway server can support many concurrent conferencing sessions in the system.

%(Chuan: I prefer a gateway server for authenticating purpose, rather than the host) A mobile user can dial in a conferencing session by calling the initiator surrogate's IP, which issues a ``heartbeat'' message including his registering information (user ID, socket address for communications, etc.) to the initiator's surrogate. 

{\em Tear-down}: When a mobile user leaves the system, its surrogate VM is released and returned to the pool of available VMs in the IaaS cloud. If the initiator of a session departs, its hosting role is handed over to another substitute surrogate in the participant list. %The session is over if fewer than $2$ concurrent participants remain. (Yu: there can be only one user, the initiator, waiting for newly-joined users)

\subsection{Routing Computation}
\label{sec:routing_calculation}

In a video conferencing session with $S$ users, there are $S$ streams, each produced by one of the mobile users, to be delivered to all the other users. For example, there are four streams in Fig.~\ref{fig:arch}, where the stream produced by one user is to be distributed to all the other three. %$n_i (i=1,2,3,4)$, is to be distributed to all other users $n_j (j\ne i)$.
 It is important to construct an efficient dissemination topology for each of the streams, to maximize the receiving rates while guaranteeing small end-to-end latencies. In addition, different participants may require different video formats/bitrates, leading to the following challenges: what is the best format/bitrate the source mobile user should send its stream at, considering needs of the receivers and bandwidth availability both among the surrogates and at the last-mile wireless links? If transcoding is necessary, at which surrogate(s) should transcoding take place along the dissemination paths, such that one transcoded stream can be useful for multiple downstream users? % as well as considering the computation capacity of the respective surrogate VMs?

We next model a mathematical optimization problem for constructing efficient dissemination topologies of all streams in a session and deciding the optimal transcoding locations. We then design efficient, fully distributed heuristic to approach the optimal solution in a dynamic system. For transcoding, we practically only consider down-sampling of a stream, {\em i.e.}, the reduction of streaming bit rate, but not the reverse, since up-sampling provides no quality improvement but consumes unnecessary bandwidth. We also focus on transcoding due to mismatched bit rates of streams of the same format, while the case of transcoding from one format to another can be readily addressed with similar efforts.

\subsubsection{Optimization Formulation}

Let graph $G=(\mathcal{S}, \mathcal{E})$ represent the network of surrogates in a session, where $\mathcal{S}$ is the set of surrogates and $\mathcal{E}$ is the set of directed connections among the surrogates. %\footnote{The graph can be a completed one with edges between each pair of surrogates, or not.}
 For each surrogate $m\in|\mathcal{S}|$, let $\hat{m}$ represent the corresponding mobile user. Let $S=|\mathcal{S}|$. Suppose $C_{ij}$ is the maximum available bandwidth on link $(i,j)\in \mathcal{E}$, and $d_{ij}$ denotes the link latency. We refer to the stream from a surrogate $m\in\mathcal{S}$ as {\em flow $m$}, with source rate $R^{(m)}_{\hat{m}}$, which is the rate of incoming stream from mobile user $\hat{m}$ to surrogate $m$, determined by the source capturing rate by the user's mobile camera and the uplink rate from the mobile user. Let $R^{(m)}_{\hat{n}}$ be the maximum acceptable bit rate of flow $m$ at mobile user $\hat{n}$, as decided by the last-mile down-link bandwidth from surrogate $n$ to $\hat{n}$, and the allocation of this down-link bandwidth among streams from different users, {\em e.g.}, if user $\hat{n}$ sizes playback windows of streams from $S-1$ other conference participants equally on its device screen, $\frac{1}{S-1}$ of the down-link bandwidth should be allocated to each stream.

As a known modeling technique, the multicast flow $m$ from surrogate $m$ to all other surrogates can be viewed as consisting of $S-1$ conceptual unicast flows \cite{zp05}%({\bf Chuan: add citation to Zongpeng's papers})
, from $m$ to each of the other surrogates, respectively. These conceptual unicast flows co-exist in the network without contending for link bandwidths, and the multicast flow rate on a link is the maximum of the rates of all the unicast flows going along this link. For ease of practical implementation, we restrict each unicast flow from $m$ to $n$ to be an integral flow along one path (but not allow fractional unicast flows along different paths), with the end-to-end rate $r^{(m)}_n$, and the multicast topology is the overlap of all the $S-1$ unicast flow paths. Let binary variable $I^{mn}_{ij}$ indicate whether the conceptual unicast flow from $m$ to $n$ traverses link $(i,j)\in \mathcal{E}$, and $c^{(m)}_{ij}$ denote the actual rate of the multicast flow $m$ on link $(i,j)$.

Let function $\varphi_n(r_1, r_2)$ give the transcoding latency at surrogate $n$, if the rate $r_1$ of an ingress flow received by $n$ is higher than the rate $r_2$ of the egress flow from $n$. $\varphi_n(r_1, r_2)=0$ if $r_1 \leq r_2$. Typical transcoding steps are to decode the source stream of rate $r_1$ to an intermediate format, and then re-encode the stream from the intermedia format to the destination rate $r_2$ \cite{transcoding}%({\bf Chuan: add citation})
. Hence, transcoding delay $\varphi_n(r_1, r_2)$ is monotonously increasing on both $r_1$ and $r_2$, and depends on computation capacity of the surrogate VM $n$: the more powerful the VM is, the faster the transcoding can be accomplished.

The quality of service in the conferencing session relies on two aspects: (i) the end-to-end latency and (ii) the flow rate received by each participant for each flow. We bound the end-to-end latency, from the time a source surrogate $m$ emits flow $m$ to the time a receiver surrogate $n$ is ready to push the stream to its corresponding mobile user, by $L^{(m)}_n$, whose value is dynamically set as in Sec.~\ref{sec:jitter_buffer}. Let $U(\frac{r^{(m)}_n}{R^{(m)}_{\hat{n}}})$ be an increasing, concave utility function on the rate of flow $m$ received by surrogate $n$, $r^{(m)}_n$. We maximize the aggregate utility of all receivers in all flows as our objective. The optimization problem is formulated in (\ref{eqn:optimization}). 

{\small
\begin{eqnarray}
          \max  \quad \sum_{m\in \mathcal{S}}\sum_{n \in \mathcal{S}, n \neq m} U (\frac{r^{(m)}_n}{R^{(m)}_{\hat{n}}})\label{eqn:optimization}
\end{eqnarray}

subject to: 

\begin{eqnarray}
		  					\sum_{i:(i,j) \in \mathcal{E}} I^{mn}_{ij} - \sum_{k:(j,k) \in \mathcal{E}} I^{mn}_{jk}=b^{mn}_j, \forall j, m,n \in \mathcal{S}, m \neq n, && \label{eqn:constr_1}\\
							I^{mn}_{ij} r^{(m)}_n\le c^{(m)}_{ij}, \forall (i,j) \in \mathcal{E}, m,n \in \mathcal{S}, m \neq n, && \label{eqn:constr_2}\\
							\sum_{m \in \mathcal{S}} c^{(m)}_{ij}\le C_{ij}, \forall (i,j) \in \mathcal{E}, && \label{eqn:constr_3}\\
							\sum_{(i,j) \in \mathcal{E}} I^{mn}_{ij}  d_{ij}
							+\sum_{(i,j)\in \mathcal{E}}\sum_{k:(j,k) \in \mathcal{E}} I^{mn}_{ij}  I^{mn}_{jk} \varphi_j (c^{(m)}_{ij}, c^{(m)}_{jk}) && \nonumber\\
							+\varphi_n (\sum_{j:(j,n) \in \mathcal{E}} I^{mn}_{jn}  c^{(m)}_{jn}, R^{(m)}_{\hat{n}}) \le L^{(m)}_n, && \nonumber\\
							\forall m,n \in \mathcal{S}, m \neq n,&& \label{eqn:constr_4}\\
	%						0 < r^{m,n} \leq R^{m,n}, \forall m,n \in \mathcal{S}, m \neq n \\
							I^{mn}_{ij} \in \{0,1\}, \forall m,n \in \mathcal{S}, m \neq n, (i,j) \in \mathcal{E},&& \label{eqn:constr_5}\\
							0\le r^{(m)}_n\le R^{(m)}_{\hat{m}},	\forall m,n \in \mathcal{S},&& \label{eqn:constr_6}\\
							0\le r^{(m)}_n\le R^{(m)}_{\hat{n}},	\forall m,n \in \mathcal{S},&& \label{eqn:constr_7}
\end{eqnarray}
}          
    where 
{\small	
\begin{equation}
b^{mn}_j =\left\{
\begin{aligned}
						-1, \quad &  j = m & \\
						1,  \quad & j = n &\\
						0,  \quad & otherwise &  \\
\end{aligned}
\right..
\nonumber
\end{equation}
}

Constraints (\ref{eqn:constr_1}) and (\ref{eqn:constr_5}) enforce a single path for the unicast flow from surrogate $m$ to $n$, and ensures flow conservation along the path. Constraint (\ref{eqn:constr_2}) implies that the unicast flow from $m$ to $n$ with rate $r^{(m)}_n$ is conceptual, ``hidden'' in the actual multicast flow $m$ with rate $c^{(m)}_{ij}$, on each link $(i,j)$. Constraint (\ref{eqn:constr_3}) requires that the overall rate of actual flows from different sources should not exceed the capacity of each link. Constraint (\ref{eqn:constr_4}) bounds the end-to-end delay along the path from source surrogate $m$ to receiver surrogate $n$, which consists of three parts: (i) the overall link delay along the path, $\sum_{(i,j) \in \mathcal{E}} I^{mn}_{ij}  d_{ij}$; (ii) the sum of potential transcoding delay at intermediate surrogates $j$'s along the path, $\sum_{(i,j)\in \mathcal{E}}\sum_{k:(j,k) \in \mathcal{E}} I^{mn}_{ij}  I^{mn}_{jk} \varphi_j (c^{(m)}_{ij}, c^{(m)}_{jk})$, where a surrogate $j$ is on the path if there exist neighboring links $(i,j)$ and $(j,k)$, such that $I^{mn}_{ij}=1$ and $I^{mn}_{jk}=1$, and a transcoding delay occurs if the flow rate on $(i,j)$, $c^{(m)}_{ij}$, is larger than the flow rate on $(j,k)$, $c^{(m)}_{jk}$; (iii) the potential transcoding delay at surrogate $n$, $\varphi_n (\sum_{j:(j,n) \in \mathcal{E}} I^{mn}_{jn}  c^{(m)}_{jn}, R^{(m)}_{\hat{n}})$, to transcode the received stream to the maximum receiving rate allowed at mobile user $\hat{n}$, if needed. Constraints (\ref{eqn:constr_6}) and (\ref{eqn:constr_7}) restrict the end-to-end rate of virtual unicast flow from surrogate $m$ to $n$ to be no larger than the maximum sending rate from mobile user $\hat{m}$ and the maximum receiving rate at mobile user $\hat{n}$.

The solutions to the optimization problem, $r^{(m)*}_n$, $c^{(m)*}_{ij}$, $I^{mn*}_{ij}$, $\forall m,n\in\mathcal{S}, n\ne m, (i,j)\in\mathcal{E}$, give us (i) the rate at which each mobile user $\hat{m}$ should send its stream to its surrogate $m$, which is the maximum of all conceptual unicast flow rates from $m$ to the other surrogates, $\max_{n\in\mathcal{S},n\ne m} r^{(m)}_n$; (ii) the delivery rate of flow $m$ along each link $(i,j)$ and hence the flow routing topology among the surrogates ($c^{(m)*}_{ij}=0$ indicates flow $m$ is not to be routed over link $(i,j)$); and (iii) where the transcoding of each flow $m$ should happen, {\em i.e.}, a surrogate $j$ where an egress flow rate $c^{(m)}_{jk}$ is smaller than the ingress rate $c^{(m)}_{ij}$ along the same conceptual unicast path, should transcode flow $m$ to the lower rate.

\subsubsection{Distributed Heuristic}

The optimization problem (\ref{eqn:optimization}) is non-convex with integer variables, thus very difficult to solve for the exact solutions. We design an efficient heuristic algorithm, as given in Alg.~\ref{alg:call_routing} and  Alg.~\ref{alg:self_evolving}, to decide flow routing, rate assignment, and transcoding in a fully distributed fashion.

We first decide a basic, feasible dissemination topology for each flow $m$, on which the end-to-end delay constraint for each receiver, constraint (\ref{eqn:constr_4}), is satisfied. Though the optimization problem (\ref{eqn:optimization}) does not restrict the topologies into trees, we seek to constraint a dissemination tree for each flow for ease of practical implementation. For conciseness, $\omega^{(m)}_n$ represents the overall latency (including both link and necessary transcoding latencies) for flow $m$ from surrogate $m$ to surrogate $n$. A shortest-path tree is constructed from surrogate $m$ to all the other surrogates, using a distributed Bellman-ford algorithm \cite{bellman} (Line 1 in Alg.~\ref{alg:call_routing}). If the overall link latency on the path from surrogates $m$ to $n$ is larger than $L^{(m)}_n$, we know that this pre-set end-to-end latency bound is by no means satisfiable, and should be adjusted to a more reasonable value (Lines 2-4). We then decide a basic, end-to-end rate of flow $m$ on this shortest path tree, from surrogate $m$ to all the other surrogates: the capacity $C_{ij}$ of each link $(i,j)$ is evenly divided by the (actual) flows generated by different surrogates, that pass through this link; the end-to-end rate of each flow $m$ is set to the rate allocated to this flow on the bottleneck link its shortest-path tree spans (Lines 5-6).

{\small
\begin{algorithm}[!t]
\caption{Flow Routing and Rate Allocation}
\label{alg:call_routing}
\begin{algorithmic}[1]
\State  Construct shortest-path trees from each surrogate $m$, $T^{(m)}$;
\If {$\exists m, n \in \mathcal{S}, \omega^{(m)}_n > L^{(m)}_n$}
	\State No feasible solution exists; \Return;
\EndIf
	\State $N_{ij} :=$ Number of dissemination trees on $(i,j)$;
	\State $\forall (a,b) \in T^{(m)}, c^{(m)}_{a,b} :=min_{k \in \mathcal{S}, (i,j) \in T^{(m)}}\{ R^{(m)}_{\hat{k}}, \frac{C_{ij}}{N_{ij}}\}$;
	\State Search for better routing paths, following Alg.~\ref{alg:self_evolving};
\end{algorithmic}
\end{algorithm}
}

%The rationale behind Alg.~\ref{alg:call_routing} is to enable a fast boot-up by constructing a shortest routing path for each flow, ignoring other co-existent flows which may contend for the link bandwidths. Once contentions occur, we equally assign the bandwidths among the transpassing streams. If even this most conservative solution, without any transcoding taken, violates the latency constraints (line 3), we can conclude no feasible solution exists. In that case, surrogates may need to be re-provisioned. Otherwise, we set further to adjust the feasible solution towards a better one iteratively, with detailed evolution mechanisms introduced in Alg.~\ref{alg:self_evolving}.  

% For ease of notation, we assume a virtual flow from surrogate $m$ to itself, {\em i.e.}, $r^{m,m}$. Obviously, $r^{m,m} = R^{0,m}$. The accumulated latency (including both transcoding and transmission latency) along the complete dissemination path for each flow $m \rightarrow n$ is noted as $l^{m,n}$, we similarly set $l^{m,m} = 0$.

Based on the basic dissemination topology, each surrogate then carries out dynamic edge and rate adjustments, in order to maximally utilize the available capacity to stream high-quality streams, without violating the latency constraints. For each flow $m$, %each surrogate maintains the end-to-end latency from the source surrogate $m$ to itself along the current tree path in $l^{(m)}_n$.
 suppose surrogate $j$ is the parent to surrogate $n$ on the current dissemination tree of flow $m$. $n$ contacts other neighboring surrogates in the flow, to discover if there is a better path from source surrogate $m$ with higher capacity %(that itself or any descent node may enjoy)
 via another parent $k$. It compares the current receiving rate $c^{(m)}_{jn}$ from $j$ with the potential receiving rate from $k$, $\min(c^{(m)}_{ik}, \bar{C}_{kn})$, where we suppose surrogate $i$ is the parent of $k$ in the current tree, and $\bar{C}_{kn}$ is the remaining available bandwidth on link $(k,n)$ (Line 2). If the potential receiving rate via $k$ is larger, $n$ needs to further evaluate the increased latency along the new path, due to changes of link latencies and potential transcoding latencies at $k$ and $n$. Only if the latency of the new path from $m$ to $n$, {\em i.e.}, $\omega^{(m)}_n$,  %, after adding potential transcoding delays at $k$
 is still within $L^{(m)}_n$, and the updated latency to each of the descent surrogates from $n$ on the tree is still within the respective delay bound, can $n$ safely change its parent from $j$ to $k$ (Lines 3-6). 
%({\bf Chuan: revise Alg.~\ref{alg:self_evolving} according to my revised text above; define necessary variable, that is not mentioned in the text, in the algorithm itself; I think it no need to discuss from virtual flow point of view, but just focus on actual multicast tree for each flow; add Line segments in Alg.~\ref{alg:self_evolving} to accompany the text}).

%Before delving into the details, some auxiliary variables are introduced. We denote $T_m$ as the dissemination path originating from surrogate $s_m$. Each node $n$ in the dissemination tree $T_m$ is assigned a two-tuples $<r_n(m), l_n(m)>$, where $r_n(m)$ represents the requested rate from surrogate $s_n$ to its upstream surrogate for flow $m \rightarrow n$, and the actual bitrate of flow $m \rightarrow n$ is no larger than requested, {\em i.e.}, $r^{m,n} \leq r_n(m)$. $l_n(m)$ is the corresponding maximal allowable latency for the stream. For instance, if $n$ is a leaf node of $T_m$, $r_n(m) = R^{m,n}$ and $l_n(m) = L^{m,n}$. Otherwise, when $n$ requests a stream, its downstream nodes' demands should also be taken into consideration. We therefore define $r_n(m) = max\{R^{m,n}, max_{j:(n,j) \in T_{m}} min \{r_j(m), C_{n,j}(m)\}\}$ and $l_n(m) = min\{L^{m,n}, min_{j:(n,j) \in T_{m}} (l_j(m) - d_{n,j} - \varphi_n(r_n(m), r_j(m)))\}$ recursively. $C_{n,j}(m)$ represents the remaining bandwidth excluded the link bandwidths allocated to flows emitting from the other surrogates $s_k, k \neq m$. 

{\small
\begin{algorithm}[!t]
\caption{Self-Evolving Route/Rate Adjustment at Surrogate $n$ in Flow $m$}
\label{alg:self_evolving}
\begin{algorithmic}[1]
	\While {$\exists (j,n) \in T^{(m)}$, $c^{(m)}_{jn} < R^{(m)}_n$}
		\If {$\exists (i,k) \in T^{(m)}$, $min\{c^{m}_{ik}, \bar{C}_{kn}\} > c^{(m)}_{jn}$}
			\State  $\Lambda := \{n\} \bigcup \{q:(n,q) \in T^{m}\}$;
			\If {$\forall p \in \Lambda, \omega^{(m)}_p \leq L^{(m)}_p$}
				\State $T^{(m)} := T^{(m)} - (j,n) + (k,n)$;
			\EndIf
		\EndIf
	\EndWhile
\end{algorithmic}
\end{algorithm}
}

\begin{figure}[h]
	\begin {center}
	\includegraphics[width=0.45\textwidth]{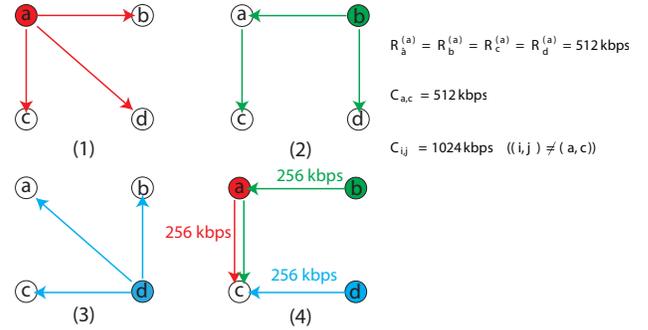}
	\caption{A simple example to illustrate the self-evolving flow routing and rate allocation algorithm.}
	\label{fig:alg1_exam}
	\end {center}
	\vspace{-3mm}
\end{figure}

We illustrate the algorithm using a simple example in Fig.~\ref{fig:alg1_exam}. There are three shortest-path trees reaching $c$, emitting from $a$, $b$ and $d$, respectively (shown in Fig.~\ref{fig:alg1_exam} (1)--(3)). Fig.~\ref{fig:alg1_exam} (4) shows the three flows reaching $c$ altogether. Suppose the only bandwidth bottleneck lies in link $(a,c)$ with a capacity of $512$ kbps, and all other links have a capacity of $1024$ kbps. %flow $a$ and $b$ contending for a total of $512$ kbps, and therefore
 The basic rate for flow $a$ and flow $b$ received by $c$ is $256$ kbps, respectively. %({\bf Chuan: draw a figure with four trees in a topology and mark link capacity on the links; revise Alg.~\ref{alg:call_routing} according to my revised text above.}).
 Then $c$ finds a better path for flow $b$ via $d$ with a higher available bandwidth of $512$ kbps, and $c$ relocates the routing path for flow $b$ after it assures that the latency constraints are not compromised.

The algorithm can be carried out in a completely decentralized fashion. Each surrogate dynamically measures the link conditions (bandwidth, delay) to its neighbouring surrogates. In our prototype implementation, latencies are measured using ``ping'' messages; bandwidth availability is estimated based on past stream transmission experience; routing path information is spread via messages exchanged between the neighboring surrogates. % We assume the link capacities are configured by mobile users, in avoidance to overweening their surrogates by bandwidth limiting for each link. We have to mention, the capacity measurements are not trivial work, especially for UDP transmissions. There have been a series of solid work in this field, which can be integrated into {\em vSkyConf} as a complement. Each surrogate will periodically inform the neighboring surrogates of the bit rates of each flows, as well as the processing power (VM configurations) it can provide. 
 It is worth noting that the adjustments at surrogates are carried out for different flows asynchronously. Such ``randomness'' allows bandwidth on a link to be allocated among different flows, rather than occupied by a few. We have carefully studied the correctness of our distributed heuristic, with theorems to show that the routing paths incur no cycles and are always feasible (guaranteeing end-to-end delay bounds). \opt{short}{Due to space limit, interested readers are referred to our technical report \cite{FullPaper}%({\bf Chuan: add citation})
  for detailed analysis.} \opt{long}{Please see the appendix for details.}

% Theorem\ref{thm:noloop} shows the adjusted dissemination paths will not fall into loops, and Theorem \ref{thm:latencybounded} guarantees the adjusted routing path is feasible.

\begin{comment}
Line $6$ in Alg.~\ref{alg:self_evolving} takes a tricky measure once the above ordinary search fails, with another simple example shown in Fig.~\ref{fig:tricky}. Surrogate $s_{n_3}$ fails to find a better routing path, since its flow rate is $400$ kbps is maximal, though a potential better path ($s_{n1}, s_{n_3}, s_{n_2}$) hides in the dark. The reason is caused by the low requested rates of $s_{n_3}$. $s_{n_2}$ will opportunistically requests $s_{n_3}$ to up-tune its requested rate to $700$ kbps, finally carving out the better routing path for $s_{n_3}$. For real-world deployment, $s_{n_2}$ may request multiple neighbouring surrogates to raise their requested rates, but only those surrogates whose original downstream surrogates' latency requirements are not compromised will help. For instance, In Fig.~\ref{alg:self_evolving}, $s_{n_3}$ will only try to up-tune its requested rates from $200$ kbps to $700$ kbps, if the latency requirement for each of its downstream surrogate is met.

\begin{figure}[h]
	\begin {center}
	\includegraphics[width=0.45\textwidth]{figures/tricky.eps}
	\caption{Opportunistic Search for Better Paths}
	\label{fig:tricky}
	\end {center}
	\vspace{-3mm}
\end{figure}

\end{comment}

\subsection{Jitter Masking}
\label{sec:jitter_buffer}

In multi-party video conferencing, a user receives multiple streams from different senders. Synchronization among different streams received at all users is crucial to users' perceived quality of experience. It is much desired that the video frames captured at all users at the same time, are played at all the recipient user devices at the same time. We design an effective buffering mechanism at the surrogates, which collaborates with the routing algorithms, for this purpose.

\begin{figure}[!t]
	\begin {center}
	\includegraphics[width=0.45\textwidth]{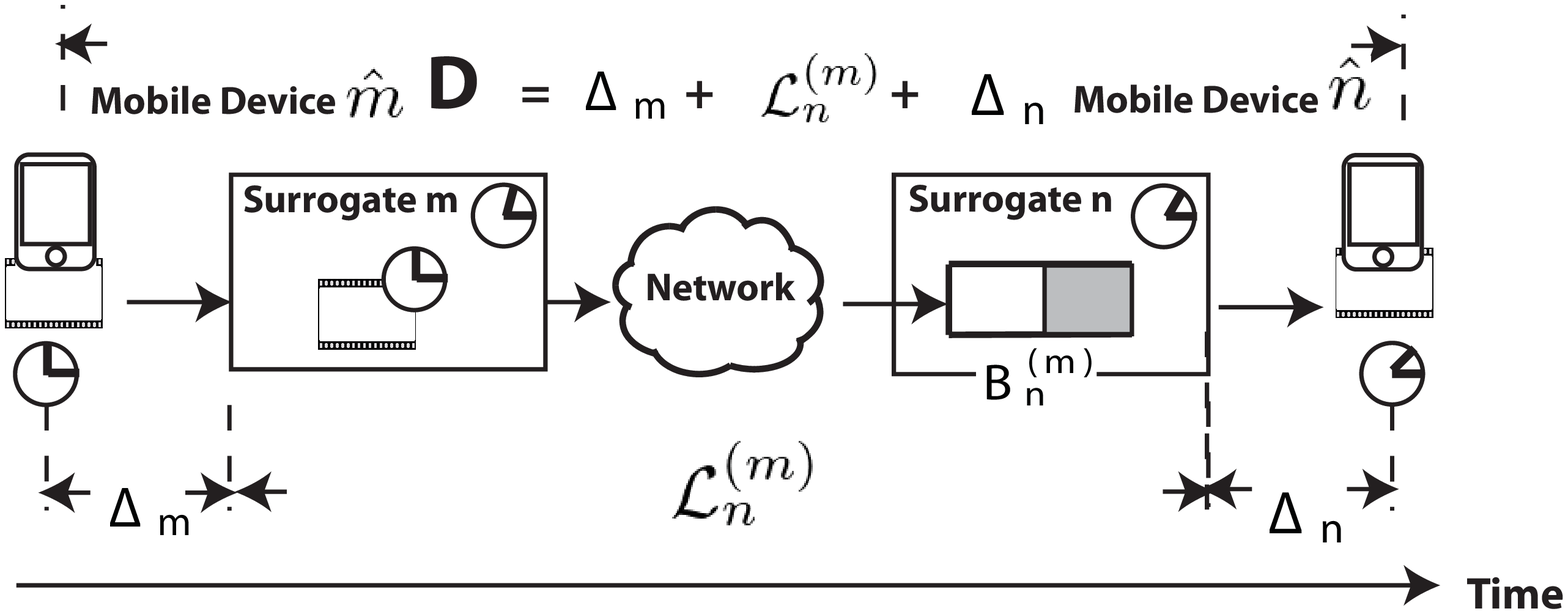}
	\caption{An illustration of the end-to-end delay for flow $m$ in {\em vSkyConf}.}
	\label{fig:fixedbuffer}
	\end {center}
	\vspace{-6mm}
\end{figure}

Surrogate $n$ maintains a buffer $B^{(m)}_n$ for each stream $m\in\mathcal{S}/\{n\}$ from each of the other surrogates. The buffer holds video packets of flow ${m}$, ready to be delivered to mobile device $n$. {\em vSkypeConf} enforces an end-to-end delay of $D$, from when a video frame is captured at one mobile device, to the time it is synchronously played at all the other mobile devices.\footnote{The value of $D$ can be set based on reasonable estimation of the maximum delay between two mobile users in the system, and should fall in the acceptable delay range for real-time communication.}  
 Let $\Delta_m$ indicate delay between mobile device $\hat{m}$ and its surrogate $m$, $\forall m\in\mathcal{S}$. For a frame in buffer $B^{(m)}_n$, which is produced at $t$ at the source $\hat{m}$, it will be pushed out from the buffer no earlier than $t+\mathcal{L}^{(m)}_n$, where $\mathcal{L}^{(m)}_n=D-\Delta_m-\Delta_n$, in order to guarantee playback of the frame at the mobile device $\hat{n}$ at $t+D$ (Fig.~\ref{fig:fixedbuffer}). Note that we seek to simply the buffering mechanism at the mobile devices in {\em vSkyConf}, while leaving the main buffering tasks to the surrogates.

If there were no jitter in the cloud, we could set the delay bound $L^{(m)}_n$ in optimization (\ref{eqn:optimization}), used to find the routing path from surrogate $m$ to surrogate $n$, to $L^{(m)}_n=\mathcal{L}^{(m)}_n$, and rest assured that the buffer will never starve. However, in a practical system, jitter may occur due to various reasons, {\em e.g.}, variation of transcoding delay at surrogates, inaccurate estimate of link delay and bandwidth when running our routing algorithm, etc. Hence, $L^{(m)}_n$ in the optimization for route selection should be set smaller than $\mathcal{L}^{(m)}_n$, in order to absorb the inaccuracy and jitter.

% Most of these delays are difficult to measure and vary over time, especially when multiple parallel transcoding tasks are taken at a single surrogate, where the the transcoding latency variations are aggravated. When delays accumulate, multiple flow streams may arrive with highly skewed latencies. To synchronize these skewed flows, those leading ones should be delayed in a buffer to wait for the lagging ones.

%{\bf Chuan: revise Fig.~\ref{fig:fixedbuffer} according to revised text;  add a time axis on the bottom and mark segments corresponding to $\Delta_m$, $\mathcal{L}^{(m)}_n$, $\Delta_n$.}

%In ideal cases where no latencies exist, the buffer is full, while in real-world situations, assuming the average accumulated latency of the flow $m \rightarrow n$ arriving at the jitter buffer is $J^{m,n}$ (different from $l^{m,n}$ defined in Sec.~\ref{sec:routing_calculation}), we can derive the average buffer load for flow $m \rightarrow n$ is $L - J^{m,n}$. 

%Starvations at any buffers may lead to glitches perceived by users, indicating latency violations in the corresponding flow. It also means, any dissemination path with a delay larger than $L$ will cause discontinuity to the corresponding flow and should be adjusted. Yet, there is no such a guarantee in real-life situations, since the jitter can be extremely large.

%From the Central Limit Theorem \cite{},
A series of solid measurement work \cite{Karam:2002:ADD:635916.635919} have shown that jitter on a network path approximately follows a normal distribution \cite{citeulike:8784}. Let $J^{(m)}_n$ be a random variable, representing the path delay from surrogate $m$ to surrogate $n$, such that $J^{(m)}_n \sim N(\mu, \sigma^2)$, %and $f (x; \mu, \sigma) = \frac{1}{\sqrt{2\pi}\sigma}e^{-\frac{(x-\mu)^2}{2\sigma^2}}$
 where $\mu$ is the mean and $\sigma$ is the standard deviation. For a normal distribution, we can derive that $99.97\%$ of the samples fall within the range of $(-\infty,\mu + 3.4\sigma)$. %Let $q^{(m)}_n$ denote the queueing delay in buffer ${B}^{(m)}_n$. 
 If we set ${L}^{(m)}_n$ to the mean $\mu$ in the path delay distribution while allowing $\mathcal{L}^{(m)}_n=\mu + 3.4\sigma$, we derive ${L}^{(m)}_n=\mathcal{L}^{(m)}_n- 3.4\sigma$. Using this ${L}^{(m)}_n$ in solving optimization (\ref{eqn:optimization}), we can make sure that $99.97\%$ of the video packets, following the path selected, can be sent out from surrogate $n$ by $\mathcal{L}^{(m)}_n$, and catch their playback deadlines at the mobile device $\hat{m}$.

%For a normal distribution, given any $P$ percentile, we can get the corresponding $\eta(P)$ easily \cite{}. For instance, $\eta(99.97\%) = \mu + 3.4\sigma$. Accordingly, we can set the maximal allowable latency, {\em i.e.}, $L^{m,n}$ as $(1-\lambda) \times L - 3.4 \sigma$ when building up the dissemination construction tree. 

% with a Probability Distribution Function (PDF) $f(\cdot)$, we still can guarantee at least a $P$ percentile distribution of $J^{m,n}$ (denoted as $\eta^{m,n}(P)$) satisfying the constraint, shown in Eqn.~\ref{eqn:jitterconstraint}. $B$ is the buffer load, sustaining a $B$ time of continuous playback.

%\begin{equation}
%	\left\{
%	\begin{array}{l}	
%			P = Prob\{J^{m,n} \leq \eta^{m,n} \}= \int^{\eta^{m,n}(P)}_{-\infty} f(x)dx \\
%			\eta^{m,n}(P) + B \leq L\\
%	\end{array}
%	\right.
%	\label{eqn:jitterconstraint}
%\end{equation}

%To avoid sudden change of jitters, the buffer load, {\em i.e.}, $B$, should be kept above a certain threshold, {\em e.g.}, $\lambda \times L$. If the maximal deviation from the average delay ($\sigma_{max}$) is known, we can estimate the maximal delay allowed, {\em i.e.}, $\mathbb{E}(J^{m,n})$, as $(1-\lambda) \times L - \sigma_{max}$, which equals exactly to $L^{m,n} $, inputs of Alg.~\ref{alg:call_routing} to construct the dissemination path to tackle the worst case.

In {\em vSkyConf}, each surrogate $n$ dynamically estimates the delay variance $\sigma$ along the path from $m$ to $n$, based on inter-packet latencies of flow $m$ it receives. It also observes the current queueing delay in buffer ${B}^{(m)}_n$, and adjusts ${L}^{(m)}_n$ used in path selection according to ${L}^{(m)}_n=\mathcal{L}^{(m)}_n- 3.4\sigma$. That is, if there are less packets in the buffer caused by larger delay variance, it tunes ${L}^{(m)}_n$ down to be more stringent on the latency requirement in the path selection; otherwise, it tunes ${L}^{(m)}_n$ up to explore paths with better bandwidths. In this way, this buffering mechanism at the surrogates collaborates with the routing algorithm, to deal with randomness in the system and inaccuracy in the computation, while maximally guaranteeing synchronized playback of all streams at all the mobile users.

\begin{comment}
For instance, the routing calculations are carried out based on both measured parameters and gossip messages between neighbouring surrogates, thus the accuracies are not $100\%$ guaranteed. What's worse, when estimating the transmission latencies in Sec.~\ref{sec:routing_calculation}, the upstream surrogates responsible for transcoding tasks are assumed not overwhelmed by their downstream surrogates, which may be too optimistic. All of these factors may lead to jitter buffer starvations in Fig.~\ref{fig:fixedbuffer}, where routing adjustments need to be taken in time. {\em vSkyConf} adopts a simple error concealment once frame packets are lost or delayed beyond the deadlines, by presenting the ``static'' last frame played instead of an annoyed ``black'' screen. More advanced and tactical error concealment will be designed in our future work.

As mentioned in Sec.~\ref{sec:data_transmissions}, inter-frames compression are still possible for the buffered $B$-length frames before delivered to the mobile clients. In that case, latency requirements are more stringent when building up the routing paths (smaller $L^{m,n}$). This can be deemed as a natural tradeoff between consumed bandwidths at mobile devices and the chat qualities perceived by the users. We will leave in-depth analysis to our future work.
\end{comment}

\section{Performance Evaluation}
\label{sec:experiments}

\subsection{Prototype Implementation and Deployment}
\label{sec:data_transmissions}

%TCP transmission solutions suit well to delay-tolerable applications, rather than time-critical ones. 
%Although RTP \cite{} has been widely adopted in real-time multimedia applications, the complexities and RTP packetization overheads ({\em e.g.}, padding after the video contents in the payloads) can not be ignored, especially for low-bitrate streams. Besides, RTP has to work in conjunction with RTCP \cite{}, whose scalability is notoriously criticized in a conferencing scenario involving a large number of participants, where the RTCP traffics may adversely occupy up to $20\%$ of the total bandwidths allocated to the session. Existing solutions seek to decrease the RTCP packets frequencies as the participant number rises, which inevitably compromises the QoS of the overall session \cite{Johnston:2009:SUS:1804456}. 

We implement a prototype of {\em vSkyConf} and deploy it in Amazon Elastic Compute Cloud (EC2), for multi-player video conferencing among users from various geographic locations. To generate reproducible experiment results, each mobile user is emulated by a machine near its assigned EC2 region (within 50 $ms$ distance), where video frames are generated at a constant rate around $1049$ kbps %every $40$ms
 ($25$ fps) from a video captured by an iSight webcam. %excluding the audio track (audio streams are usually transmitted separately \cite{Firestone:2007:VVC:1406748} and irrelevant to this work). 
 Surrogates are provisioned  from ``ap-southeast-1a'' region (Singapore) for Hong Kong users, ``eu-west-1a'' (Ireland) for European users, ``us-west-1b'' (California) and ``us-east-1a'' (Virginia) for users in west US and east US, respectively. To showcase the adaptability and self-healing capability of our system against abrupt network fluctuations, we emulate  dynamic environments by manually injecting jitters on the links between surrogates via Dummynet \cite{Dummynet}. We implement an application-layer packet controller to limit the link capacities between surrogates, into the range of [128,1050] Kbps.  %, in terms of the transmission bandwidths for {\em vSkyConf} packet payload by calculating away the overheads of headers.
 Both uplink and downlink bandwidths of each emulated mobile user are within the range of $[1.5, 2]$ Mbps, the same as those on regular 3G cellular connections. We apply the concave function $\log(x)$ as the utility function in our routing computation. The latencies between surrogates are the actual delays between Amazon EC2 instances. The transcoding latencies, are pre-evaluated on the VM instances and used in our routing computation, for transcoding from $768$kbps to $256$kbps, from $768$kbps to $128$kbps, from $256$kbps to $128$kbps, respectively, which are all the cases for transcoding under our setup. On each of our emulated mobile clients, the stream from one of the other conference participants is displayed in a large screen (corresponding to a maximal acceptable streaming rate of $768$ kbps), and streams from other participants are displayed using smaller screens (corresponding to maximal acceptable streaming rates of $128$ kbps or $256$ kbps). Besides, a fixed $400$ ms end-to-end delay ($D$ in Sec.~\ref{sec:jitter_buffer}) is configured, and the buffer for each flow at each surrogate is set to a size corresponding to $400$ms stream playback.

\begin{figure}[h]
	\begin {center}
	\includegraphics[width=0.45\textwidth]{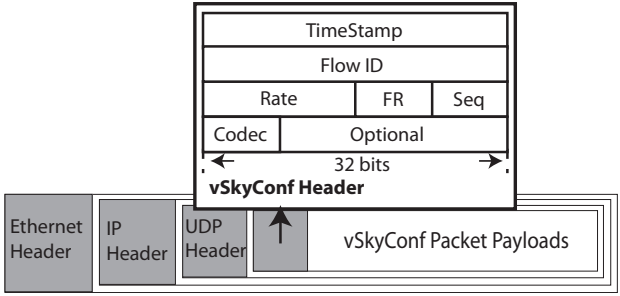}
	\vspace{-2mm}
	\caption{{\em vSkyConf} packet header.}
	\label{fig:mediapackets}
	\end {center}
	\vspace{-6mm}
\end{figure}

A light-weighted stream transmission protocol among surrogates is implemented based on UDP. %To sustain stable video flows between each pair of users, we apply a constant bit rate (CBR) transcoding style instead of variable bit rate (VBR), to avoid disruptive quality degradation caused by bit rate surges. 
 The packet header has $13$-octet mandatory part with $3$ octets for future extensions, as shown in Fig.~\ref{fig:mediapackets}. The ``TimeStamp'' (4 bytes) represents the moment when the packet is generated. Before attaching the current time to a packet, the surrogate should add in the first-mile latency between the mobile user and itself. The ``Flow ID'' field (4 bytes) indicates who generates the packet, by including the IP address of the source surrogate. ``Rate'' (2 bytes) represents the bit rate (kbps) of the stream encapsulated in the packet, ``FR'' (1 byte) stands for the frame rates (fps)  and ``Seq'' (1 byte) is the sequence number of the packet in the flow. %for a frame starting from $1$,
  There are a total number of $\lceil \frac{Rate}{FR \times P_{max}} \rceil$ packets for each video frame, where $P_{max}$ represents the maximal payload length of a {\em vSkyConf} packet, which is chosen to be $512$ bytes in our implementation. %to avoid accidental packet truncations by the underlying network.
 A frame is lost if any packet of this frame is lost. ``Codec'' field (1 byte) indicates the codec of the stream for transcoding reference at the surrogates. 

%{\em vSkyConf} packets only exist in {\em vSkyConf} overlay networks established by surrogates, the packet format at the mobile side is irrelevant, which enables {\em vSkyConf} flexible deployments on a vast range of platforms, or even as a transportation protocol for ordinary real-time multi-party communications.

%Different from audio-only streams which can be sampled around $10$ ms (G.711 \cite{}), video frames are generated much more slowly where $25$ frames per second is already qualified enough for smooth visual chats. However, video conferencing poses extremely stringent latency requirements, thus {\em vSkyConf} only takes intra-frame compressions (transcoding), since inter-frame compressions require buffering multiple frames and hence incur intolerable latencies. In sec.~\ref{sec:jitter_buffer}, we discuss further improvements for inter-frame compressions, which is left as our future work.

\subsection{Adaptive Flow Rates at vSkyConf Clients}
\label{sec:large_test}

We test a video conferencing session among $10$ participants: 5 from Hong Kong, 1 from Europe, 2 from US West and 2 from US East, respectively. Random jitters up to $150$ ms are imposed on the links between the surrogates in Europe and in Hong Kong. %, which are also susceptible to network instability in real-life measurement \cite{EC2_benchmark}.

\begin{figure}[h]
	\begin {center}
	\includegraphics[width=0.45\textwidth]{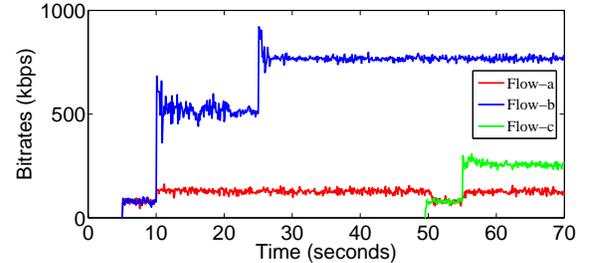}
	\vspace{-2mm}
	\caption{Flow rates at the initiator's surrogate.}
	\label{fig:normal_flow_rates}
	\end {center}
	\vspace{-6mm}
\end{figure}

As a potential bottleneck for scalability, the surrogate for the session initiator is responsible to maintain the session by handling the ``heartbeat'' messages and periodically broadcasting the user lists to the other surrogates in the session. We therefore investigate the conferencing performance at the initiator's surrogate: if its performance is satisfactory, then the performance at the other surrogates should be even better. Fig.~\ref{fig:normal_flow_rates} illustrates the flow rates for streams from $3$ among the other $9$ conference participants (since plotting $9$ curves in a figure would make it less readable). ``Flow-b'' is the flow from the European user, configured to be displayed at the main large screen at the initiator mobile user (corresponding to a maximum streaming rate of $768$kbps); ``flow-a'' and ``flow-c'' are to be displayed at smaller screens at the initiator (corresponding to a maximum streaming rate of $128$kbps and $256$kbps, respectively), coming from Hong Kong and US west, respectively, with the latter joining the session at a later time. We can see that both ``flow-a'' and ``flow-b'' go through a ``fast'' start stage, when the basic stream dissemination topology is being constructed (as introduced in Sec.~\ref{sec:routing_calculation}), and then evolve towards their maximal acceptable rates. ``flow-a'' achieves its maximal rate quickly, while ``flow-b'' sticks to the link between ``eu-west-1a'' and ``ap-southeast-1a'' before redirected to a better routing path, as affected by the injected jitters along the link. After ``flow-c'' joins the session around $47$ seconds, the flow rate of ``flow-a'' drops a bit before ``flow-c'' adjusts its routing path shortly. This is caused by the link contention between the new ``flow-c'' and the existing ``flow-a''. 

\begin{figure}[h]
	\begin {center}
	\includegraphics[width=0.45\textwidth]{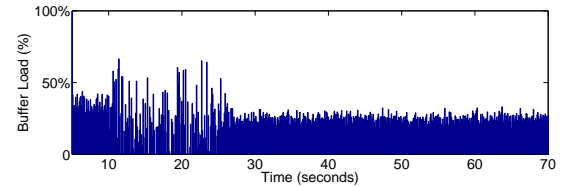}
	\vspace{-2mm}
	\caption{Load of flow-b's buffer at the initiator's surrogate.}
	\label{fig:exp_buffer}
	\end {center}
	\vspace{-6mm}
\end{figure}

Fig.~\ref{fig:exp_buffer} presents the load in the jitter buffer for ``flow-b'' at the initiator's surrogate, where we see that the buffering level varies significantly when ``flow-b'' takes a path going through the link between ``eu-west-1a'' and ``ap-southeast-1a'', due to jitters long the link. This causes $L^{(m)}_n$ to be tuned down and hence the end-to-end latency constraint in optimization (\ref{eqn:optimization}) for ``flow-b'' is violated. The algorithm then redirects ``flow-b'' through a better path via the ``us-west-1b'' region, which leads to a more stable buffering level later on.% after the initiator receives its ``path broadcast'' message notifying a better path. %We believe the up-soaring rate peak right before a path adjustment is caused by the ``in-the-air'' packets before a relocation command takes effect.

\begin{figure}[h]
	\begin {center}
	\includegraphics[width=0.45\textwidth]{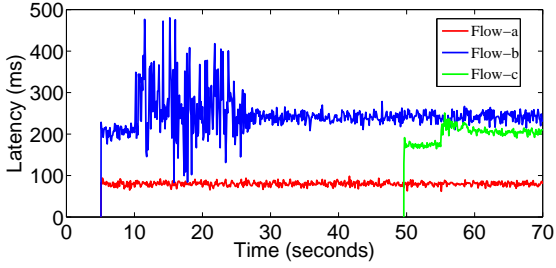}
	\vspace{-2mm}
	\caption{Flow latencies at the initiator's surrogate.}
	\label{fig:normal_flow_latency}
	\end {center}
	\vspace{-6mm}
\end{figure}

Fig.~\ref{fig:normal_flow_latency} shows the corresponding latency of each flow, from the corresponding source surrogate to the initiator's surrogate. We observe that latencies only vary slightly whenever the routing paths are adjusted, and can well meet the end-to-end latency required ($400$ms). The latency of ``flow-b'' varies more significantly, due to the manually imposed jitters on the link between ``eu-west-1a'' and ``ap-southeast-1a'', before the algorithm redirects the flow to a better path. %Though ``flow-b'' is finally redirected to the US west region (California), the average latency does not improve, because we found the latency between ``eu-west-1a'' and ``ap-southeast-1a'' approximately equals to the sum of latencies of  link from ``us-west-1b'' to ``ap-southeast-1a'' and link from ``eu-west-1a'' to ``us-west-1b''.

All the above results show that the streaming rates are promptly adaptive to the network conditions among the surrogates, and the stream playback at the initiator is quite smooth with stable end-to-end delays up to the requirements.

\subsection{Performance Comparison with a Unicast Solution}

\begin{figure}[h]
	\begin {center}
	\includegraphics[width=0.45\textwidth]{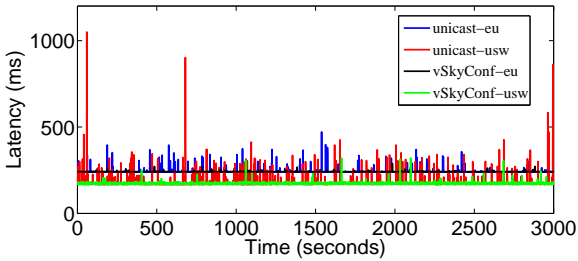}
	\vspace{-2mm}
	\caption{End-to-end latency experienced at the Hong Kong user.}
	\label{fig:latency_comparison}
	\end {center}
	\vspace{-6mm}
\end{figure}

We next evaluate the performance of {\em vSkyConf} against a unicast scheme typically applied by peer-to-peer video conferencing solutions, where each flow is directly transmitted from the source to the destination via the cellular network (which is via the Internet in our emulated experiments). To conduct a fair comparison, we establish a 3-user video conferencing session, since the uplink bandwidth limits the conference size in a unicast scheme. We emulate a $50$-minute long conferencing session with one user coming from each of the regions, Hong Kong, Europe, and west US, respectively. Other experimental settings are the same as used in Sec.~\ref{sec:large_test}, except that no emulated jitters are imposed on the link from ``eu-west-1a'' to ``ap-southeast-1a''. Fig.~\ref{fig:latency_comparison} shows the perceived end-to-end latencies of the two flows received at the user in Hong Kong when it employs the unicast solution or {\em vSkyConf}, where ``eu'' stands for Europe and ``usw'' stands for west US. We can see that the end-to-end latency achieved with {\em vSkyConf} is generally smaller, and much more stable than that achieved by the unicast solution. 

\begin{figure}[h]
	\begin {center}
	\includegraphics[width=0.45\textwidth]{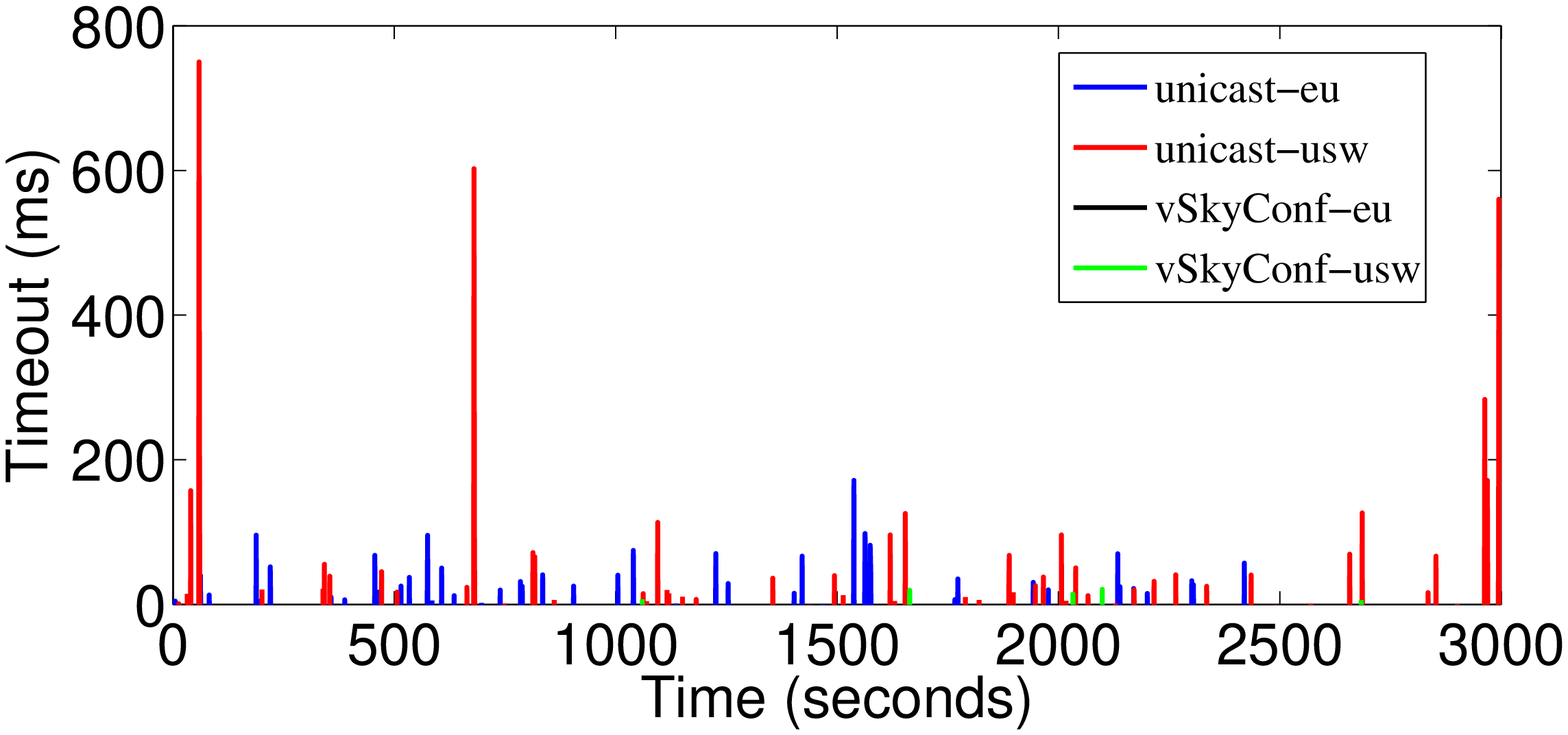}
	\vspace{-2mm}
	\caption{Packet time-out delay at the Hong Kong user.}
	\label{fig:latency_timeout}
	\end {center}
	\vspace{-6mm}
\end{figure}

Fig.~\ref{fig:latency_timeout} compares the streaming smoothness between the two solutions, by evaluating the amount of time-out delay incurred during the streaming of each flow, due to late packets received after their playback deadlines. The x value indicates the occurrence time of packet time-out and the y value indicates the packet delay beyond the respective deadline. Again, much fewer packets arrive after their playback deadlines in {\em vSkyConf}, verifying the smooth stream playback experienced by {\em vSkyConf} users. This shows that our cloud-assisted design is very suitable to achieve high-quality video conferencing among multiple mobile participants.
 
\section{Conclusion and Future Work}
\label{sec:conclusions}

This paper presents {\em vSkyConf}, a cloud-assisted mobile video conferencing system, designed to fundamentally improve the quality and scale of multi-party mobile video conferencing. In {\em vSkyConf}, a virtual machine in a cloud infrastructure is employed as the proxy for each mobile user, to send and to receive conferencing streams, and to transcode the streams into proper formats/rates. We design a fully decentralized, efficient algorithm to decide the best paths of stream dissemination and the most suitable surrogates for video transcoding along the paths, and tailor a buffering mechanism on each surrogate to cooperate with optimal stream distribution. These designs guarantee bounded, small end-to-end latencies and smooth stream playback at the mobile devices. We have implemented {\em vSkyConf} based on Amazon EC2 and verified the excellent performance of our design, as compared to the widely adopted unicast solutions. As future work, we seek to test {\em vSkyConf} under more dynamic settings.
\opt{long}{\appendix
\label{sec:appendix}

\begin {theorem}
		No cycles exist in the adjusted dissemination tree.
		\begin{IEEEproof}
			The initial feasible solution contains no cycles, since it is a shortest-path tree $T^{(m)}$. Suppose the cycle happens when a surrogate $i$ adjust its upstream surrogate from $i'$ to $k$,  {\em i.e.}, $\cdots, i', i, j, \cdots, k', k, \cdots$. The down-sampling transcoding mechanisms applied in our algorithm guarantees that $c^{(m)}_{i'i} \geq c^{(m)}_{ij} \geq \cdots \geq c^{(m)}_{k'k}$. According to Alg.~\ref{alg:self_evolving}, $i$ can only adjust its upstream surrogate to $k$ when $c^{(m)}_{k'k} > c^{(m)}_{i'i}$, which contradicts.
		\end{IEEEproof}
		\label{thm:noloop}
\end {theorem}

Alg. \ref{alg:self_evolving} sketches the core logic, independent of any specific implementations. Our prototype, {\em i.e.}, {\em vSkyconf}, enables gossip-like message exchanges between neighbouring surrogates to facilitate the construction of dissemination trees. 

Each surrogate $m$ maintains two key tables, {\em i.e.}, Candidate Upstream Surrogate Table ($CUSTab_m$) and Downstream Surrogate Table ($DSTab_m$), respectively. CUSTab keeps track of the possible paths for each flow, from which the surrogate can choose one immediately once routing adjustment is needed. DSTab keeps track of all the down stream surrogates for each flow to which the surrogate should relay after necessary transcoding. Based on constantly updated DSTab, each surrogate is associated with a metric pair for each flow, {\em i.e.}, $<requested-rate, maximal-delay>$. $requested-rate$ represents the rate the surrogate should request from the upstream surrogate, while $maximal-delay$ represents the maximal delay when choosing a path and can help filter out those unqualified upstream surrogates. Let $\alpha^{(m)}_i$ represent the requested rate for flow $m$ at surrogate $i$, and $\beta^{(m)}_i$ represent the maximal delay for flow $m$ when surrogate $i$ chooses a path. Both of them can be defined recursively as Eqn.~\ref{eqn:requestrate} and Fig.~\ref{eqn:maxdelay}.

{\small	
\begin{equation}
\alpha^{(m)}_i =\left\{
\begin{aligned}
						min\{R^{(m)}_{\hat{i}},\bar{C}_{i'i}\} \quad  (DSTab_i = \phi )& \\
						min\{max_{j \in DSTab_i}\{R^{(m)}_{\hat{i}}, \alpha^{(m)}_j\},\bar{C}_{i'i}\} & \\
						(DSTab_i \neq \phi) &  \\
\end{aligned}
\right.,
\label{eqn:requestrate}
\end{equation}
}
where $i'$ is the upstream surrogate of $i$ for flow $m$.

{\small
\begin{equation}
\beta^{(m)}_i =\left\{
\begin{aligned}
  L^{(m)}_i \quad (DSTab_i = \phi) & \\
  min_{j \in DSTab_i} \{L^{(m)}_i, \beta^{(m)}_j - d_{ij}-\varphi_i(c^{(m)}_{i'i}, c^{(m)}_{ij})\} & \\
   (DSTab_i \neq \phi) & \\
\end{aligned}
\right..
\label{eqn:maxdelay}
\end{equation}
}

Each surrogate $i$ issues ``Path Broadcast'' messages, informing its neighbouring surrogates of the bit rates of any flows it can offer, as shown in Fig.~\ref{fig:gossip}. ``Rate'' represents the current rate for flow labelled by ``Flow ID''. ``MaxRate'' represents the requested rate defined above. ``Latency'' represents the actual latency from the source surrogate to the current surrogate, {\em i.e.}, $\omega^{(m)}_i$. ``VM configuration'' represents the VM instance type of the surrogate. Once receiving a ``Path Broadcast'' message for flow $m$ from surrogate $j$, a surrogate $i$ will only have it recorded into $CUSTab_i$ if the latency $\omega^{(m)}_i$ is no larger than $\beta^{(m)}_i$, which can be estimated as $\omega^{(m)}_j + d_{ji} + \varphi_j (\alpha^{(m)}_j, \alpha^{(m)}_i)$.

\begin{figure}[h]
	\begin {center}
	\includegraphics[width=0.4\textwidth]{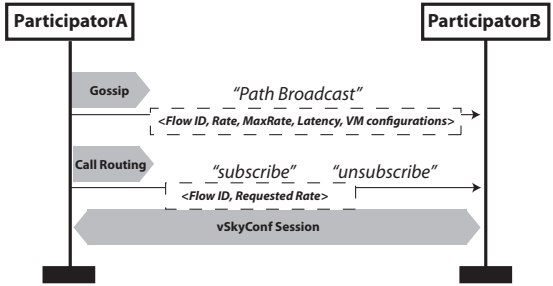}
	\caption{\em Gossip-like message communication.}
	\label{fig:gossip}
	\end {center}
	\vspace{-3mm}
\end{figure}

\begin{lemma}
	For any adjacent surrogates $i$ and $j$ in a dissemination tree $T^{(m)}$ ($(i,j) \in T^{(m)}$), $\varphi_i(\alpha^{(m)}_i, \alpha^{(m)}_j) \geq \varphi_i(c^{(m)}_{i'i}, c^{(m)}_{ij})$, where $i'$ is the upstream surrogate of $i$ for flow $m$.
	\begin{IEEEproof}
		we know $c^{(m)}_{i'i} \leq \alpha^{(m)}_i$, $c^{(m)}_{ij} \leq \alpha^{(m)}_j$ and $\alpha^{(m)}_i \geq \alpha^{(m)}_j$ according to the definition. If $\alpha^{(m)}_i > \alpha^{(m)}_j$, $\varphi_i(\alpha^{(m)}_i, \alpha^{(m)}_j) > \varphi_i(c^{(m)}_{i'i}, c^{(m)}_{ij})$ since the transcoding latency $\varphi(\cdot, \cdot)$ is monotonously increasing on both the input and output bit rates (Clarified in Sec.~\ref{sec:design}); Otherwise, $\alpha^{(m)}_i = \alpha^{(m)}_j$, suppose $c^{(m)}_{i'i} \neq c^{(m)}_{ij}$ ({\em i.e.}, $c^{(m)}_{i'i} > c^{(m)}_{ij}$ under a down-sampling only mechanism), we can derive $\bar{C}_{ij} \geq \alpha^{(m)}_j = \alpha^{(m)}_i \geq c^{(m)}_{i'i} > c^{(m)}_{ij}$. It means the actual bit rate of flow $m$ perceived by surrogate $j$ is lower than both the requested bit rate and the remaining link bandwidth, which doesn't conform to Alg.~\ref{alg:self_evolving}. So $c^{(m)}_{i'i} = c^{(m)}_{ij}$, and $\varphi_i(\alpha^{(m)}_i, \alpha^{(m)}_j) = \varphi_i(c^{(m)}_{i'i}, c^{(m)}_{ij}) = 0$. 
	
	To sum up, in either case, $\varphi_i(\alpha^{(m)}_i, \alpha^{(m)}_j) \geq \varphi_i(c^{(m)}_{i'i}, c^{(m)}_{ij})$.
	\end{IEEEproof}
	\label{lemma:transcodinglatency}
\end{lemma}

\begin {theorem}
		The adjusted solution guarantees the latency bounds for impacted surrogates.
		\begin{IEEEproof}
			Once the dissemination tree $T^{(m)}$ is adjusted by redirecting surrogate $i$ to a new upstream surrogate under the condition $\omega^{(m)}_i \leq \beta^{(m)}_i$, the potentially affected nodes can only be in the sub tree of $T^{(m)}$ rooted from $i$. Suppose the latency perceived by surrogate $j$ in the sub tree violates the latency constraint, {\em i.e.}, $\omega^{(m)}_j > L^{(m)}_j$. We could find a path from $i$ to $j$. Assume the upstream surrogate of $j$ is $k$, the corresponding latency is $\omega^{(m)}_j = \omega^{(m)}_k + d_{kj} + \varphi_k(c^{(m)}_{k'k}, c^{(m)}_{kj})$, where $k'$ is the upstream surrogate of $k$ for flow $m$. Since $L^{(m)}_j \geq \beta^{(m)}_j$ according to the definition, $\varphi_k(c^{(m)}_{k'k}, c^{(m)}_{kj}) \leq \varphi_k(\alpha^{(m)}_k, \alpha^{(m)}_j)$ (guaranteed by Lemma \ref{lemma:transcodinglatency}), we have, $\omega^{(m)}_k = \omega^{(m)}_j - d_{kj} - \varphi_k(\omega^{(m)}_k, \omega^{(m)}_j) > L^{(m)}_j - d_{kj} - \varphi_k(\alpha^{(m)}_k, \alpha^{(m)}_j) \geq \beta^{(m)}_j - d_{kj} - \varphi_k(\alpha^{(m)}_k, \alpha^{(m)}_j) \geq \beta^{(m)}_k$. Similarly, along the path, we can easily derive $\omega^{(m)}_i > \beta^{(m)}_i$, which contradicts.
		\end{IEEEproof}
		\label{thm:latencybounded}
\end {theorem}

}

\bibliographystyle{IEEEtran}
\bibliography{main}

\end{document}